\documentclass[structabstract]{aa}
\usepackage[dvips]{graphicx}
\usepackage{amssymb}
\usepackage{xcolor}    %\textcolor{red}{some text}
\usepackage[normalem]{ulem}
\usepackage{graphicx}
\usepackage{txfonts}
\usepackage[sort]{natbib}
\bibpunct{(}{)}{;}{a}{}{,}

\def\etal{{\rm et al. }}

\def\kpc{{h^{-1} \rm kpc}}
\def\kms{{\rm km\, s^{-1}}}
\def\aap{{\em A}\&{\em A}}

\def\aj{{\em AJ}}
\def\apj{{\em ApJ}}
\def\apjl{{\em ApJ}}
\def\apjs{{\em ApJS}}
\def\araa{{\em ARA}\&{\em A}}

\def\mn{{\em MNRAS}}
\def\mnras{{\em MNRAS}}
\def\nat{{\em Nature}}
\def\nature{{\em Nature}}
\def\pasp{{\em PASP}}

\begin{document}

   \titlerunning{The impact of bars and interactions on optically selected AGNs}
   \authorrunning{Alonso et al.}

   \title{The impact of bars and interactions on optically selected AGNs in spiral galaxies}

%   \subtitle{}

   \author{Sol Alonso\inst{1},
          Georgina Coldwell\inst{1},
          Fernanda Duplancic\inst{1},
          Valeria Mesa\inst{2}
                  \and
          Diego G. Lambas\inst{3}
          }

   \institute{Departamento de Geof\'{i}sica y Astronom\'{i}a, CONICET, Facultad de Ciencias Exactas, F\'{i}sicas y Naturales, Universidad Nacional de San Juan, Av. Ignacio de la Roza 590 (O), J5402DCS, Rivadavia, San Juan, Argentina\\
              \email{solalonsog@gmail.com.ar}
         \and
            Instituto Argentino de Nivolog\'{i}a, Glaciolog\'{i}a y Ciencias Ambientales, IANIGLA-CONICET, Parque Gral San Mart\'{i}n, CC 330, CP 5500, Mendoza, Argentina 
         \and  
Observatorio Astron\'omico de C\'ordoba, OAC, CONICET, Universidad Nacional de C\'ordoba, Laprida 854,
X5000BGR, C\'ordoba, Argentina
             }
             
   \date{Received xxx; accepted xxx}

   \abstract
  % context heading (optional)
  % {} leave it empty if necessary  
   {}
     % aims heading (mandatory)
   {With the aim of performing a suitable comparison of the internal process of galactic bars with respect to the external effect of interactions on driving gas toward the inner most region of the galaxies, we explored and compared the efficiency of both mechanisms on central nuclear activity  in optically selected active galactic nuclei (AGNs) in spiral galaxies.}
 % methods heading (mandatory)
   {We selected homogeneous samples of barred AGNs and active objects residing in pair systems, derived from the Sloan Digital Sky Survey (SDSS).
In order to carry out a reliable comparison of both samples (AGNs in barred hosts in isolation and in galaxy pairs), we selected spiral AGN galaxies with similar distributions of redshift, magnitude, stellar mass, color and stellar age population from both catalogs. 
With the goal of providing an appropriate quantification of the influence of strong bars and interactions on nuclear activity, we also constructed a suitable control sample of unbarred spiral AGNs without a companion and with similar host properties to the other two samples.}
  % results heading (mandatory)
   {We found that barred optically selected AGNs show an excess of nuclear activity (as derived from the $Lum[OIII]$) and accretion rate onto a central black hole ($\cal R$) with respect to AGNs in pairs. 
In addition, both samples show an excess of high values of $Lum[OIII]$ and $\cal R$ with respect to unbarred AGNs in the control sample.

We also found that the fractions of AGNs with powerful nuclear activity and high accretion rates increase toward more massive hosts with bluer colors and younger stellar populations. Moreover, AGNs with bars exhibit a higher fraction of galaxies with powerful $Lum[OIII]$ and efficient $\cal R$ with respect to AGN galaxies inhabiting pair systems, in bins of different galaxy properties.

Regarding AGNs belonging to pair systems, we found that the central nuclear activity is remarkably dependent on the galaxy pair companion features. The $Lum[OIII]$ for AGNs in pairs is clearly enhanced when the galaxy companion exhibits a bright and more massive host with high metallicity, blue color, efficient star formation activity and young stellar population.

The results of this work reveal an important capacity of both mechanisms, bars and interactions, to transport material towards the galaxy central regions. 
In this context, it should also  be noted that the internal process of the bar is more efficient at improving the central nuclear activity in AGN objects than that corresponding to the external mechanism of the galaxy-galaxy interactions.}
   {}


   \maketitle 
%
%________________________________________________________________

\section{Introduction}

The most accepted hypothesis surrounding the origin or active galactic nuclei (AGNs) proposes that they arise from accretion of material onto a central massive black hole triggering nuclear activity.
It is widely known that the main fueling mechanisms of the central engine in galaxies are related to dynamical perturbations transporting gas to the inner most central regions \citep{LB69,Rees84}. 
In this sense, several authors agree that galactic bars and galaxy mergers/interactions are usually considered the two principal processes for torquing material to the centers of active galaxies (e.g., Mihos \& Hernquist 1996; Combes 2003; Alonso et al. 2007, Di Matteo et al. 2005; Alonso et al. 2013, 2014).   

Bars play a fundamental role in the dynamical evolution of their host galaxies and can also affect several properties of spirals on relatively short timescales.
In this context, bar perturbations can modify star formation activity, stellar populations, colors, chemical composition and even galactic structure \citep{atha83,buta96,comb93,cheu13,martin95,robi17,vera16}, promoting evolution of their host galaxies  (Ellison et al. 2011a, Zhou et al. 2015).
Moreover, the gas infall produced
by bars toward the innermost regions of galaxies is a mechanism that may efficiently trigger nuclear activity in the central zone of AGN galaxies \citep{cor03,comb93,ju18}.
Different studies based on numerical simulations show a loss in galaxy angular momentum produced by interactions between gas clouds and the edges of the bar, driving a flow of material toward the central regions of barred galaxies \citep{SBF90}. \\

Following this line, \cite{pet18} analyzed different mass models of spiral galaxies under the influence of tidal interactions finding no strong correlation between
bar length or pattern speed and the interaction strength. However, these authors show that interactions slightly accelerate bar formation in some models. On the other hand, there is a slower  disk-dominated  rotation-curve  model  likely  due  to  interactions of  gas  clumps. In agreement with this point, \cite{zana18} carried out a study of external versus internal bursts of bar formation, finding a strong dependence on the mass of the disk.\\

Furthermore, the ``bars within bars'' scenario states that an external bar transports material over distances of a few parsecs. In this region an internal secondary bar produces gravitational instability in the accumulated gas, enabling flow toward regions near the massive black hole \citep{SBF89}. 
Supporting this theory, several authors using different observational techniques have observed secondary bars inside strong external bars \citep{emse01,malk98,laine02,caro02,mac2000}. In this context, in a recent study,  \cite{du17} suggest that these short bars, generally embedded in large-scale  bars, are an  important  mechanism  for  driving  gas  inflow, feeding the central black hole. Nevertheless, this mechanism becomes unstable, and inner bars are destroyed when the
black hole mass  grows  to $\sim0.1\%$ of the total stellar mass. This event slows down, or even stops, the growth of a central black hole.

There is clear observational evidence that bars enhance central nuclear activity compared to non-barred spiral galaxies: 
 \cite{oh12} found that bars produce an increment in the  central activity of blue galaxies with low black 
hole masses from a sample of barred late-type AGNs. Furthermore, Alonso et al. (2013, hereafter A13) show that isolated barred AGN galaxies display a higher fraction of powerful nuclear activity, in comparison with a suitable control sample of unbarred AGNs with similar distributions of redshift, magnitude, morphology and local environment.
They also found that barred AGN galaxies show an excess of objects with high accretion rates in comparison 
to unbarred ones.
In addition, from the analysis of barred AGN spiral galaxies inhabiting groups and clusters, Alonso et al. (2014) found that the increment of nuclear activity produced by bar perturbations is also notable in barred active galaxies located in higher-density environments. Recently, \cite{gal15}, using a sample of disk galaxies from Sloan Digital Sky Survey (SDSS) and Galaxy Zoo 2, found a higher fraction of barred AGNs than of star-forming barred galaxies, although the central black hole accretion rate shows no dependence on the presence of a bar. \cite{cheu15} deepened this study towards high redshifts concluding that large-scale bars cannot be considered the dominant fueling mechanism for supermassive black hole growth; see also \cite{gou17}.

Galaxy interactions can be an effective mechanism to modify different host galaxy properties, mainly by triggering star formation  \citep{alo06,alo12,barton,Lam03, Lam12,kenni,mesa}, and also affect the galaxy stellar mass function \citep{gama}. The presence of a close galaxy companion drives a clear enhancement in galaxy morphological asymmetries, and this effect is statistically significant up to projected separations of at least 50 $\kpc$ \citep{patton16}.
The physical processes behind galaxy-galaxy interactions have been explained by theoretical and numerical analyses (e.g., Martinet 1995; Toomre \& Toomre 1972; Barnes \& Hernquist 1992, 1996; Mihos \& Hernquist 1996), showing that collisional disruption, material dissipation, and gas inflows are  produced by the tidal torques generated during near encounters.
In addition to feeding star formation, these material inflows could also feed a central black hole and increase nuclear activity (Sanders et al. 1988). 
The performance of this process depends on the gas reservoir and the particular internal characteristics of galaxies involved in the interaction. 

Based on observational evidence, the connection between mergers/interactions and nuclear activity is fairly well accepted.  
In this sense, several studies have found clear clues of interactions in luminous quasar hosts 
(e.g., Canalizo \& Stockton 2001; Bennert et al. 2008; Urrutia, Lacy \& Becker 2008; Ramos Almeida et al. 2011; Bessiere et al. 2012; Urrutia et al. 2012). 
In addition, different analyses have found a clear increase of the nuclear activity in less-luminous AGNs with tidal interaction features or distorted morphologies with respect to non-interacting AGN galaxies (e.g., Koss et al. 2010, 2012; 
Ellison et al. 2011a, 2013; Silverman et al. 2011; Sabater et al. 2013). 
In particular, Alonso et. al (2007) performed a statistical analysis of nuclear activity comparing AGN galaxies in 1607 close 
pair systems to AGNs without companions. They found that the nuclear activity of active galaxies with strong interaction 
features is significantly larger than for AGNs in an isolated environment. 
The accretion rate also shows that AGNs in merging pairs are actively feeding their central black holes. \cite{mesa} constructed a sample of spiral galaxy pairs from SDSS, and classified them according to the spiral arms' rotation pattern, detecting an increment in the nuclear activity in systems whose spiral arms rotate in opposing directions. 
More recently, Sabater et al. (2015) found that galaxy interactions affect AGN activity in an indirect way, by influencing the central gas supply.
In this sense, these studies provide obvious clues about AGN fueling and its link with galactic mergers and interactions. In addition, \cite{bar17} asserts that the enhancements in specific star formation rates (SFRs) are positively correlated with enhanced AGN luminosity, suggesting that both values are mutually triggered by the merger events, the latter being significantly greater than the former. 

An interesting approach is to study and compare the role of bars and galaxy interactions in feeding central black holes.
Motivated by this, we analyze the effect of the internal process of bars in comparison with the external mechanism of galaxy 
interactions on the central nuclear activity in AGN galaxies. 
For this purpose, using data from the Sloan Digital Sky Survey, we obtain large and homogeneous samples of barred AGN 
galaxies and AGNs in pair systems required to derive a direct and consistent comparison of these two mechanisms. 
The conclusions of these studies will allow us to expand on what is currently known about the governing mechanism of the induction of radial gas inflow to galactic centers.

This paper is structured as follows. Section 2 describes the procedure used to construct the samples of barred AGN galaxies,  AGNs in pairs, and control AGN galaxies.
In Section 3, we study the effects of bars and galaxy interactions on nuclear activity and the relation with the host galaxy properties.
Section 4 explores the role of the galaxy pair companion in feeding central black holes, and in Sect. 5 we summarize our main conclusions.  
The cosmology adopted here is  
$\Omega = 0.3$, $\Omega_{\lambda} = 0.7$, and $H_0 = 100~ \kms \rm Mpc$.

%__________________________________________________________________

\section{Samples}

This work is based on photometric and spectroscopic data selected from the 
Sloan Digital Sky Survey Data Release 7 (SDSS-DR7) galaxy catalog (Abazajian et al. 2009). SDSS-DR7 comprises 11663 square degrees of sky imaged in five wave bands (u, g, r, i and z) between 3543\AA\  and 9134\AA, and provides imaging data for 357 million objects. 
In this release, the main galaxy sample is essentially a magnitude limited spectroscopic sample with $r_{lim}<17.77$ (Petrosian magnitude) covering a redshift range $0<z<0.25$, with a median redshift of 0.1 (Strauss et al. 2002). Several galaxy physical properties have been used in this study. 
The procedures to derive these properties are described by Brinchmann et al. (2004), Tremonti et al. (2004) and Blanton et al. (2005).
These data are available from  MPA/JHU%
\footnote{http://www.mpa-garching.mpg.de/SDSS/DR7/ %
} and the NYU% 
\footnote{http://sdss.physics.nyu.edu/vagc/ %
}, including 
 emission-line fluxes, stellar masses, SFR indexes, gas-phase metallicities, and so on.
 As an indicator of the age of stellar populations, we adopted the spectral index $D_n(4000)$. 
This parameter represents an important effect in the spectra of old stars, that occurring at 4000\AA\   \citep{kauff03}, and arises by an accumulation of a large number of spectral lines in a narrow region of the spectrum.
In this work, we have adopted the \cite{balo99} definition.

For the optically selected AGNs from SDSS (AGN-SDSS) we used the standard diagnostic
diagram proposed by Baldwin et al. (1981, hereafter BPT). This diagram allows 
for the separation of type II AGNs from normal star-forming galaxies using emission-line ratios. These emission-lines were corrected for optical reddening using the Balmer decrement and the obscuration curve \citep{calzetti00}. 
The signal-to-noise ratio (S/N) was calculated with the emission-line flux errors adjusted according to the uncertainties 
suggested by the MPA/JHU catalog\footnote{http://www.mpa-garching.mpg.de/SDSS/DR7/raw$\_$data.html}.
Furthermore, we used only galaxies with an S/N $> 2$ for all lines intervening
in the diagnostic diagram used to distinguish AGNs from
HII galaxies in view of the fact that the adjusted uncertainties almost duplicated the original errors. Hence, considering the relation between spectral
lines [OIII]λ5007, $H\beta$, $[NII]\lambda 6583$, and $H\alpha$ within the BPT diagram,
we followed the \cite{kauff03} criterion to select type II AGN as those with

\begin{equation}
log_{10}([OIII]/H\beta) > 0.61/(log_{10}([NII/H\alpha]) − 0.05) + 1.3
.\end{equation}

\subsection{Barred AGN galaxy catalog}
 
For the analysis of this paper, we use the catalog of barred AGN spiral galaxies obtained in our previous work (A13).
The procedure performed in A13 to construct this catalog is summarized below. 

In order to obtain spiral active hosts, we cross-correlated 
the AGN-SDSS galaxies with the spiral objects obtained from the Galaxy Zoo \textbf{1}\footnote{http://www.galaxyzoo.org/} \citep{zoo, zoo2}.
This catalog comprises a morphological classification of nearly 900000 galaxies drawn from the SDSS.
This survey is contributed by hundreds of thousands of volunteers,  however due to the large number of classifiers, it becomes complex to maintain a list of unified criteria and a reliable means of classification.
They  define different categories (i.e., elliptical, spiral, merger, uncertain, etc.) and give the fraction of votes in each category.
In this study, we selected galaxies that were classified as spiral galaxies by the Galaxy Zoo team with a fraction of votes $>0.6$, and in this way
a low fraction of galaxies with nonspiral morphological types could be included. 
We also restricted AGN spiral galaxies to redshifts of $z<0.1$, $g-$band petrosian apparent magnitude ($g-$mag) brighter than 16.5, and axial ratio $b/a>0.4$. These restrictions lead to the selection of  face-on bright galaxies which favor the classification based on the naked-eye detection.
In addition, the galaxies with these constraints were classified as barred or unbarred by visual inspection using the $g$ + $r$ + $i$ combined color images, from on-line SDSS Image Tool% 
\footnote{http://skyserver.sdss.org/dr7/en/tools/chart/list.asp %
}, finding 1927 barred and 4638 unbarred active spiral galaxies.

In order to maintain homogeneous criteria, bar detection by visual inspection was performed by just one of the authors % 
\footnote{Sol Alonso %
}. The reliability of the classification was addressed by comparison with the classification of a subsample of barred galaxies by another author % 
\footnote{Georgina Coldwell %
}. 
This procedure allows us to quantify the uncertainty in the classification derived through visual inspection. Although this type of classification could be seen to be relatively subjective, the level of agreement in the bar detection in spiral galaxies was very high (95$\%$  overlap between the  samples of both classifiers).

In addition, to assess the accuracy of our selection criteria, we cross-correlated our sample with barred AGN galaxies
taken from the catalog of the Nair \& Abraham (2010).
This catalog was obtained from the fourth data release (DR4) of SDSS and detected a total of 454 barred active galaxies with the similar constraints in $g-$mag and redshift to those of our sample.
The catalog of the barred active galaxies constructed for this work was obtained from SDSS DR7 which covers a greater area than DR4 and represents a homogeneous sample selected with unified criteria and a reliable method of classification.   
We find the level of agreement to be more than acceptable, with 96.5$\%$  overlap between both samples.

Furthermore, in order to derive unbiased results regarding the  effects of bars on the central nuclear activity, we derived a sample of AGNs hosted by isolated barred galaxies, requiring that any neighboring galaxy within a region
of 500 kpc $h^{-1}$ in projected separation and $\Delta V<$ 1000 km $s^{-1}$ in relative velocity must be fainter than the barred active galaxy. With this criterion, we obtained a sample of 1530 isolated barred AGN spiral galaxies 
(see A13 for more details).

\subsection{Catalog of AGNs hosted by spiral galaxies in pairs }

In order to obtain AGNs hosted by spiral galaxies in pairs, 
we built a galaxy pair catalog, requiring members to have projected separations of $r_{p}< 100 \rm \,kpc \,h^{-1}$
and relative radial velocities of $\Delta V< 500 \rm \,km \,s^{-1}$ within $z<0.1$. 
The number of pairs satisfying these criteria is 30437.
With the aim to analyze in detail the effect of interactions in spiral active galaxies, we identified spiral AGNs in pairs by cross-correlating the total galaxy pair catalog with AGN-SDSS galaxies and with spirals identified in the Galaxy Zoo catalog. From these cross-correlations, we obtain a sample of 6906 pairs, where one of the galaxy members of the pair system is an AGN spiral galaxy.
 We also restricted AGN spirals belonging to pairs with a $g-$band magnitude brighter than 16.5, and thus obtained 2970 pairs.  

Furthermore, bars can themselves be the result of galaxy-galaxy interactions \citep{Moet17,kaza08,nogu87}.   
Therefore, during close encounters between galaxies there is an important redistribution of mass and a strong gravitational tidal torque that could form a central bar. In this sense, we also explored the frequency of bars in AGN galaxies in pair systems, finding that 6.5$\%$ (446 objects) of the active objects in pairs are barred spiral galaxies.
%We noticed a low percentage of galaxies with bars in interaction in our sample (or has a pair companion within 100 kpc).  
For a sample of 294 galaxies with bars taken from the catalog of Nair
\& Abraham (2010), Ellison et al. (2011a) do not find spiral galaxies with bars belonging to close pair systems within $r_{p}< 30 \rm \,kpc$.
With the aim of analyzing the effect of bars and interactions on the central nuclear activity independently of one another, we extracted these barred spiral active galaxies from the sample of AGNs in pairs. 

In order to make a suitable comparison between the internal processes of bars and the external effect of the mergers/interactions with respect to the inflow of gas toward the central region of the  active galaxies, we selected AGN spirals in both samples with similar distributions of redshift, $g-$mag, absolute $r-$band magnitude (hereafter luminosity or $M_r$) and stellar mass (see panels (a), (b), (c) and (d) in Fig.~\ref{histcont}).
We also considered AGN galaxies in pair systems and barred AGN hosts with similar color ($M_u-M_r$) and stellar age population ($D_n(4000)$) distributions, with the aim to obtain similar host galaxy properties in both samples (see panels (e) and (f) in Fig.~\ref{histcont}).
With these restrictions, the final barred galaxy catalog (hereafter \textsc{barred-agn}) contains 1060 barred AGN galaxies and the catalog of AGN spiral galaxies in pairs (hereafter \textsc{agn in pairs}) is composed of 2890 AGN spiral galaxies in pair systems. 
Eighty percent of the galaxies of \textsc{agn in pairs} is composed of non-AGN and AGN galaxy pairs systems and 20$\%$ are pairs composed of two active objects.

\subsection{Control sample}

To provide a reliable quantification of the impact of bars and interactions on the central nuclear activity, we also constructed an appropriate control sample of isolated unbarred spiral AGNs, i.e., AGN galaxies without a pair companion, that share similar host galaxy properties to those of the samples described above. 

From the sample of 4638 unbarred AGN spiral galaxies previously classified (see Section 2.1), we selected isolated objects using a suitable isolation criterion defined for the barred AGN sample, and we also filtered out objects with a pair galaxy companion within $r_{p}< 100 \rm \,kpc \,h^{-1}$ and $\Delta V< 500 \rm \,km \,s^{-1}$. With the aim to check the isolated criterion, in A13 we analyzed the distribution of the local density environment in the barred and unbarred galaxies. 
With this purpose, for both isolated samples (barred and unbarred) 
we defined a projected local density parameter, $\Sigma_5$.
This parameter was calculated through the projected distance $d$ 
to the fifth nearest neighbor, $\Sigma_5 = 5/(\pi d^2)$.
The neighbor galaxies were chosen to have luminosities  $M_r < -20.5$  \citep{balo04} and with a radial velocity difference of less than 1000 km $s^{-1}$.
The $\Sigma_5$ distributions are plotted in Fig. 1 (panel (e)) of the our previous paper (A13), showing a trend towards low $\Sigma_5$ values what implies low-density local environments. 

\begin{figure}
\centering
\includegraphics[width=80mm,height=130mm ]{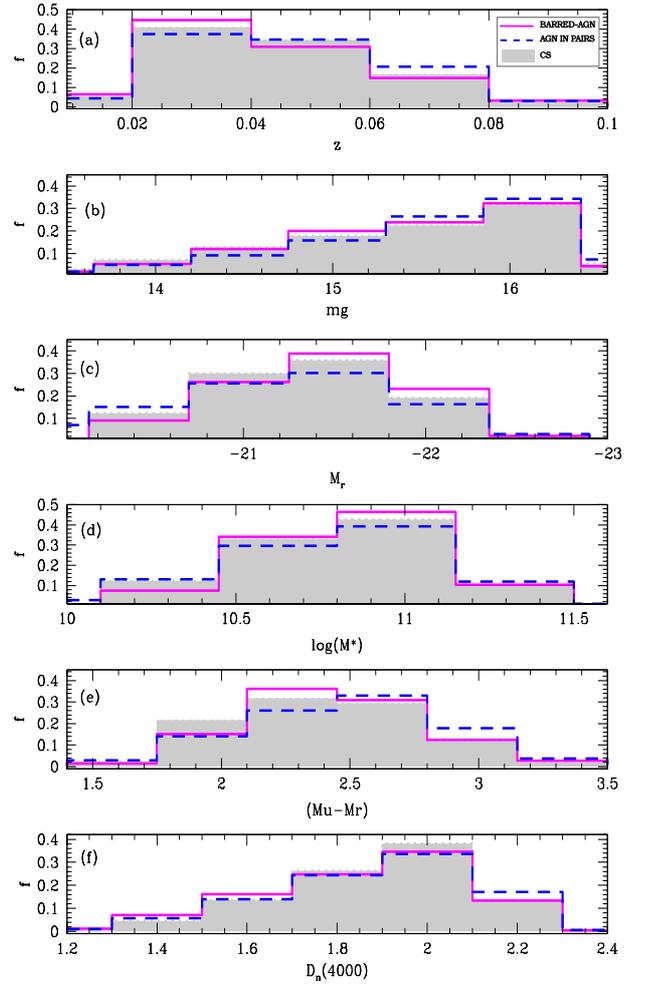}
\caption{Normalized distributions of redshift, $z$, $g-$band apparent magnitude, $g-$mag, luminosities, $M_r$,
 stellar mass, $log(M*)$, color index, ($M_u-M_r$), and stellar age population, $D_n(4000)$, 
(panels $a$, $b$, $c$, $d$, $e$ and $f$, respectively),
for barred AGN galaxies (solid lines), AGN galaxies in pair systems (dashed lines)
and unbarred spiral AGN objects (full surfaces).
}
\label{histcont}
\end{figure}

Subsequently, we defined a sample of control galaxies using a Monte Carlo algorithm
that selects galaxies in the unbarred isolated sample with similar distributions of redshift, $g-$mag, luminosities, stellar mass, $M_u-M_r$ colors and stellar age population to those of the barred AGN host sample. In addition, these distributions are similar to those of the \textsc{AGN in pairs}. The dotted lines in Fig.~\ref{histcont} show the distributions of these  parameters for the control sample. 
We performed a Kolmogorov Smirnov (KS) test between the distributions of the control sample and the distributions of barred AGN galaxies. We also carry out a KS test between the distributions of the control sample and the distributions of \textsc{AGN in pairs}.
 From this test we obtain a $p$ value that
represents the probability that a value of the KS statistic will be
equal to or more extreme than the observed value, if the null hypothesis holds. In all cases, we obtained $p$$>$0.05 for the null hypothesis
that the samples were drawn from the same distributions.

Finally, the control sample (hereafter \textsc{CS}) contains 1242 isolated unbarred spiral AGN galaxies without a pair companion. The details and numbers of the different AGN galaxy catalogs obtained during this work are listed in Table 1. Figure 2 shows images of typical examples of AGN galaxies selected from the different samples.
The procedure followed to construct these catalogs ensures that they will have the same host properties, and, consequently, can be used to estimate a clear difference between  bars and interactions, in comparison with regular spiral AGNs, unveiling the effect of  both mechanisms on the central nuclear activity.

\begin{figure}
\centering
\includegraphics[width=80mm,height=125mm ]{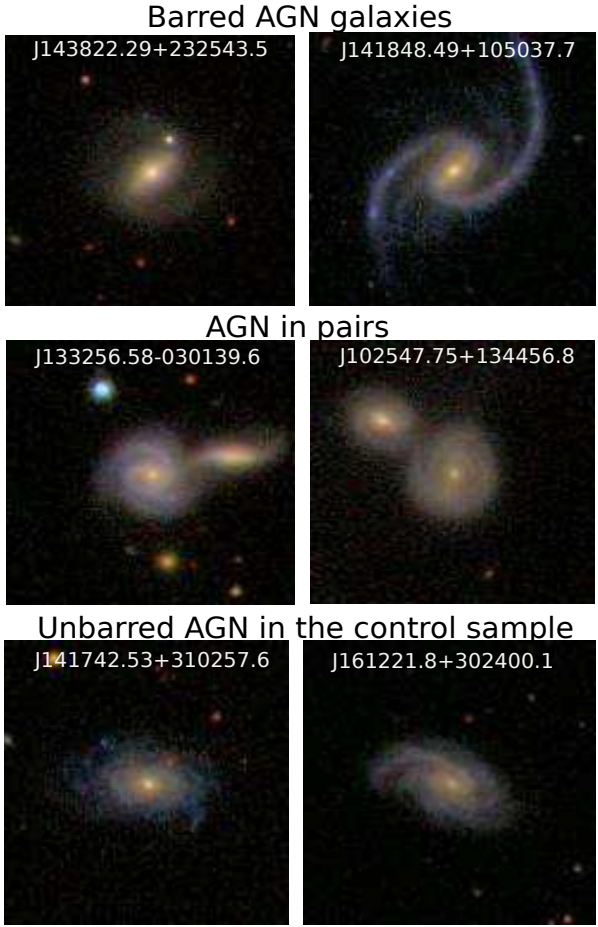}
\caption{Images of typical examples of active galaxies in our samples. The scale of the images is $\approx$ 1$'$.5 x 1$'$.5. 
}
\label{ejemplos}
\end{figure}

\begin{table}
\center
\caption{Catalogs obtained for this work with similar host galaxy properties (see Fig.1) .}
\begin{tabular}{|c c c| }
\hline
Sample & Criterion &  Number   \\
\hline
\hline
\textsc{barred-agn} & isolated barred AGN spiral galaxies   & 1060   \\
\textsc{agn in pairs} & AGN spiral galaxies in pair systems   & 2890  \\
\textsc{CS} & isolated unbarred AGN spiral galaxies & 1242     \\
\hline
\end{tabular}
{\small}
\end{table}

%-----------------------------------------------------------------------------
\section{Nuclear activity: effect of bars and interactions}

In this section, we perform a comparative analysis of the impact of bars and mergers/interactions on the nuclear activity driven by feeding of the central black hole. 
To this aim, we focus on the dust-corrected luminosity of the [OIII]$\lambda$5007 line, $Lum[OIII]$, as a tracer of the AGN nuclear  activity.  
The [OIII] line is one of the strongest narrow emission lines in optically obscured AGNs and has very low contamination by contributions of star formation in the host galaxy. 
The $Lum[OIII]$ estimator has been widely studied by several authors in different analyses \citep{mul94, kauff03, heck04, heck05, brinch04}. 
In addition, \cite{kauff03} found low contamination from star formation lines in the $Lum[OIII]$ for high-metallicity host galaxies. In this context, most of the AGNs in our samples have stellar masses $M^* >10^{10}$ $M_{\sun}$ (see panel (d) in Fig. 1), therefore, following of mass-metallicity relation \citep{tremonti04}, the metallicities are expected to be high, and therefore with low levels of contamination from star formation lines.

\begin{figure}
\centering
\includegraphics[width=80mm,height=80mm]{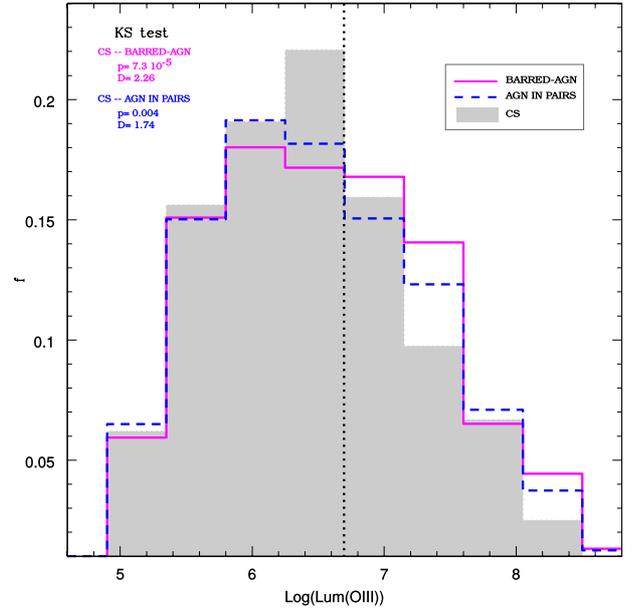}
\caption{Distribution of $log(Lum[OIII])$ for barred AGN galaxies (solid line), AGNs in pair systems (dashed line) 
and unbarred AGNs in the control sample (full surfaces).
The D and $p$-values of the KS test between both samples (\textsc{barred-agn}, \textsc{agn in pairs}) and the \textsc{CS} is inset in the upper-left corner of the figure.
The dotted vertical line represents the excess of $Lum[OIII]$ for the \textsc{barred-agn}.
}
\label{histLO}
\end{figure}

Figure~\ref{histLO} shows the distributions of $Lum[OIII]$ for barred AGN  objects and AGNs in galaxy pairs.
We also plot unbarred active galaxies in the control sample.
We note a tendency of the barred AGN galaxies to have larger nuclear activity values with respect to AGNs in pair systems. In addition, both samples show an excess of high $Lum[OIII]$ with respect to the control sample.
The Kolmogorov-Smirnov statistics allow us to quantify the difference between these distributions with a significance of 99,99\%. The D and $p$-values of the KS test between both samples (\textsc{barred-agn}, \textsc{agn in pairs}) and the \textsc{CS} in $Lum[OIII]$ are presented in the legend of the Fig.~\ref{histLO}.

We divide the samples into low- and high-luminosity subsamples, considering the threshold $Lum[OIII] = 10^{6.4} L_{\odot}$ as a suitable limit which approximately
 corresponds to the mean $Lum[OIII]$ of the control sample. This threshold has also been used in our previous works (A13; Alonso et al. 2014).  
Furthermore, we consider AGN galaxies with $Lum[OIII] > 10^{7.0} L_{\odot}$ as extremely powerful active galaxies. 
In a similar direction, Coldwell et al. (2009) considered weak AGNs to be those with $Lum[OIII] < 10^{6.45}$ and powerful AGNs to be those with $Lum[OIII] > 10^{7.07} L_{\odot}$.
The same limit for strong AGNs was also defined by Kauffmann et al. (2003) for a sample of 22623 optically selected AGNs from the BPT diagram.  
These authors also found that quasars (QSO) and strong type II AGNs, both selected with  $Lum[OIII] > 10^{7.0} L_{\odot}$, present similar young stellar content and establish that a young population is associated with all types of AGNs with strong $[OIII]$ emission. 
However, Zakamska et al. (2003) present a catalog of type II quasars from the SDSS, selected based on their optical emission lines considering $Lum[OIII] > 3x10^{8.0} L_{\odot}$. A similar threshold in $[OIII]$ luminosity was used  by Reyes et al. (2008) to select type II QSOs.  
Table 2 quantifies the percentage of AGN galaxies with high luminosity and extremely powerful nuclear activity in the three samples studied in this work.

With the aim to analyze the accretion strength of the central black holes, we have also calculated the accretion rate 
parameter, $\cal R$=log($Lum[OIII]$/$M_{BH})$, \citep{heck04}.
Using the correlation between the black hole mass, $M_{BH}$, and the
bulge velocity dispersion, $\sigma_*$ (Tremaine et al. 2004),
we first estimated $M_{BH}$ for the three samples analyzed in this paper. 

\begin{equation}
logM_{BH} = \alpha + \beta log(\sigma_*/ 200).
\end{equation}

\cite{gra08} found that the central velocity dispersion is enhanced by the stellar motion along a bar, and therefore 
the $M_{BH} - \sigma_*$ relation is different for barred and unbarred galaxies.
We adopted ($\alpha$, $\beta$) = (7.67 $\pm$ 0.115, 4.08 $\pm$ 0.751) for barred active galaxies 
and ($\alpha$, $\beta$) = (8.19 $\pm$ 0.087, 4.21 $\pm$ 0.446) for AGNs in pairs and for unbarred AGN host in the control sample \citep{gulte09}. 
For this analysis, we filtered out galaxies with $\sigma_* > 70 km s^{-1}$, because the instrumental resolution
of SDSS spectra is $\sigma_* \approx$ 60 to 70 $km s^{-1}$.

We expect galaxies with low central velocity dispersions are likely to show  late-type morphologies with very small bulge masses (like those spiral galaxies shown in Fig. 2). 
Recently, Simmons, Smethurst \& Lintott (2017) found a sample of disk-dominated galaxies capable of growing supermassive black holes with very efficient rates.
This finding reflects the theoretical prediction by Martin et al. (2018) who showed how disk-dominated galaxies with small-bulge to total-stellar-mass  ratios (B/T $<$ 0.1) can have black hole masses approximately one to two times more massive than one would expect given their bulge masses. 
In this context, we found that $\sim$ 15$\%$ of the AGN galaxies in our samples have $\sigma_* <$ 100 $km s^{-1}$, which could have massive black holes with substantial growth rates. 

\begin{figure}
\centering
\includegraphics[width=70mm,height=70mm ]{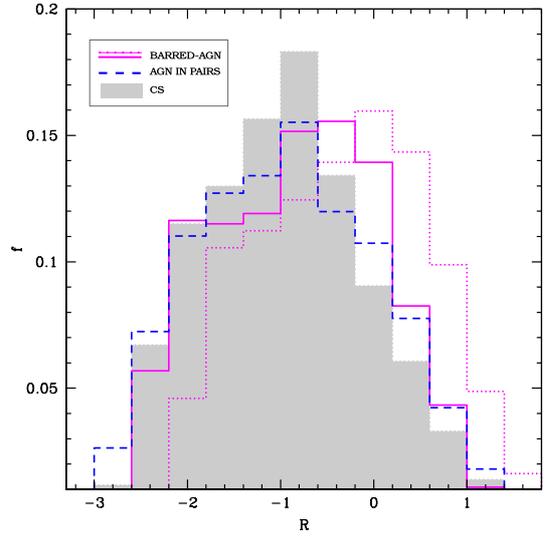}
\caption{ Distribution of accretion rate 
parameter, $\cal R$=log($Lum[OIII]$/$M_{BH})$, for barred AGN galaxies (dotted line), AGNs in pair systems (dashed line), 
and unbarred AGNs in the control sample (full surfaces). The solid line represents the $\cal R$ distribution for barred AGN galaxies, 
 using the same parameters, $\alpha$ and $\beta$, as for the unbarred galaxies to calculate $M_{BH}$ (see Eq. 2).}
\label{histR}
\end{figure}

Figure \ref{histR} shows the $\cal R$ distributions for barred AGN galaxies, AGNs in pair systems, 
and unbarred AGNs in the control sample.
Clearly, active galaxies in pairs show an excess of high accretion rate values with respect to the control sample. Furthermore, barred AGNs exhibit higher $\cal R$ values with respect to the other samples.  
We also calculated the accretion rates of barred AGN hosts using the same parameters, $\alpha$ and $\beta$, 
that we used to obtain $M_{BH}$ and $\cal R$ for the other AGN samples (represented by a solid line in Fig.\ref{histR}). Although in this  case, the signal is less significant, barred active galaxies also show an excess of accretion rate compared to the other active samples.
The difference of these distributions was also quantified by the Kolmogorov-Smirnov statistics (with confidence of 99.8\%). 
From these distributions, we defined $\cal R=$-0.6 as a value from which the excess of accretion rate for the barred AGN sample is  noticeable with respect to the other samples (see Table 2).

In a similar way, several authors have studied the accretion strength of the central black holes in different AGN samples through the accretion rate 
parameter \citep[e.g.,][]{heck04,alo07,alo13,coldwell14,gal15}. 
Estimations of the mean accretion rates of AGNs in
pair systems in comparison with isolated active galaxies indicate that, at a given
luminosity or stellar mass content, AGNs in merging pairs have more active black holes than other AGNs \citep{alo07}.
Moreover, barred AGN hosts show an excess of objects with high accretion rate values with respect to unbarred active galaxies (A13).
On the other hand, \cite{gal15} studied the strength of AGNs using the $\cal R$ parameter from a sample of 353 barred Seyfert galaxies with respect to 328 unbarred Seyferts, showing that barred AGNs do not exhibit stronger accretion than unbarred AGNs.
The discrepancy in these results could be due to the different  methods used to  select the AGN samples. 
These authors chose Seyfert AGN galaxies with lower stellar masses and z$<$0.05; however, they tested and compared $\cal R$ distributions for barred and unbarred galaxies considering a similar AGN selection (adding LINER and composite galaxies) with similar mass and redshift ranges to those of A13, finding good agreement in the results.

The findings shown in this section indicate that bar perturbations are a more efficient mechanism to drive radial gas inflows towards the central zone, improving the nuclear activity and the accretion rate of black holes in AGN galaxies, with respect to the process of mergers and interactions between galaxies.

\begin{table*}
\center
\caption{Percentages of AGN galaxies with high-luminosity, extremely powerful nuclear activity, and high accretion rate values in the three samples studied in this work: \textsc{barred-agn}, \textsc{agn in pairs} and \textsc{CS}.}
\begin{tabular}{|c c c c | }
\hline
Restrictions & $Lum[OIII] > 10^{6.4} L_{\odot}$ &  $Lum[OIII] > 10^{7.0} L_{\odot}$ & $\cal R$ $>$-0.6 \\
\hline
\hline
$\%$ of \textsc{barred-agn}  &   55.2$\%$$\pm$0.6 &   34.3$\%$$\pm$1.1 &  43.3$\%$$\pm$0.7 \\
$\%$ of \textsc{agn in pairs}  &   52.3$\%$$\pm$0.3 &   29.5$\%$$\pm$0.7 &  37.2$\%$$\pm$0.5 \\
$\%$ of \textsc{CS}  &   48.7$\%$$\pm$0.5 &   24.3$\%$$\pm$0.8 &  33.5$\%$$\pm$0.6 \\
\hline
\end{tabular}
{\small}
\end{table*}

%-------------------------------------------------------------------------------------
\subsection{Influence of host galaxy properties on nuclear activity}

In this section we explore the impact of bars and interactions on the black hole activity and accretion rate with 
respect to the host galaxy properties.

We highlight that our AGN sample is optically selected. The different adopted AGN selection criteria at different wavelengths could lead to the selection of distinct host galaxies with morphologies, accretion rates, and environments varied (e.g., Tasse et al. 2008; Hickox et al. 2009; Smolcic 2009; Best \& Heckman 2012). In this context, Ellison et al. (2016) compared the host galaxy SFRs of different AGN samples, finding that infrared (IR), optical , and radio-selected AGN form a sequence in SFR that is consistent with an evolutionary scenario in which gas supply triggers enhanced star formation in an obscured AGN phase.
Furthermore, by using numerical simulations, Hopkins et al (2005) demonstrated that the same processes that trigger the nuclear activity generate a darkening of the AGNs over most of their lifespan, which can be observed in the optical during a short period of time.

Similarly, Hickox et al. (2009) proposed an evolutionary picture showing a sequence with decreasing star formation and accretion rate for AGN samples obtained from observations in IR, X-ray, and radio frequencies. 
In addition, from the study of the powerful (high-excitation) and the weak (low-excitation) radio AGNs, Smolcic (2009) found that these two samples represent the earlier and later stages, respectively, in the formation of massive galaxies, suggesting that it could be linked to different states of the AGN feedback.

These results imply a dependence between the AGN selection criteria from different wavelengths and the AGN evolutionary state, which is reflected in host galaxy properties and the accretion rate of the central black hole.
We stress the fact that the samples analyzed in this work are selected optically using homogeneous criteria, so that the analysis of host properties and nuclear activity reflect only the effect of bars and interactions.

\begin{figure*}
\centering
\includegraphics[width=175mm,height=75mm ]{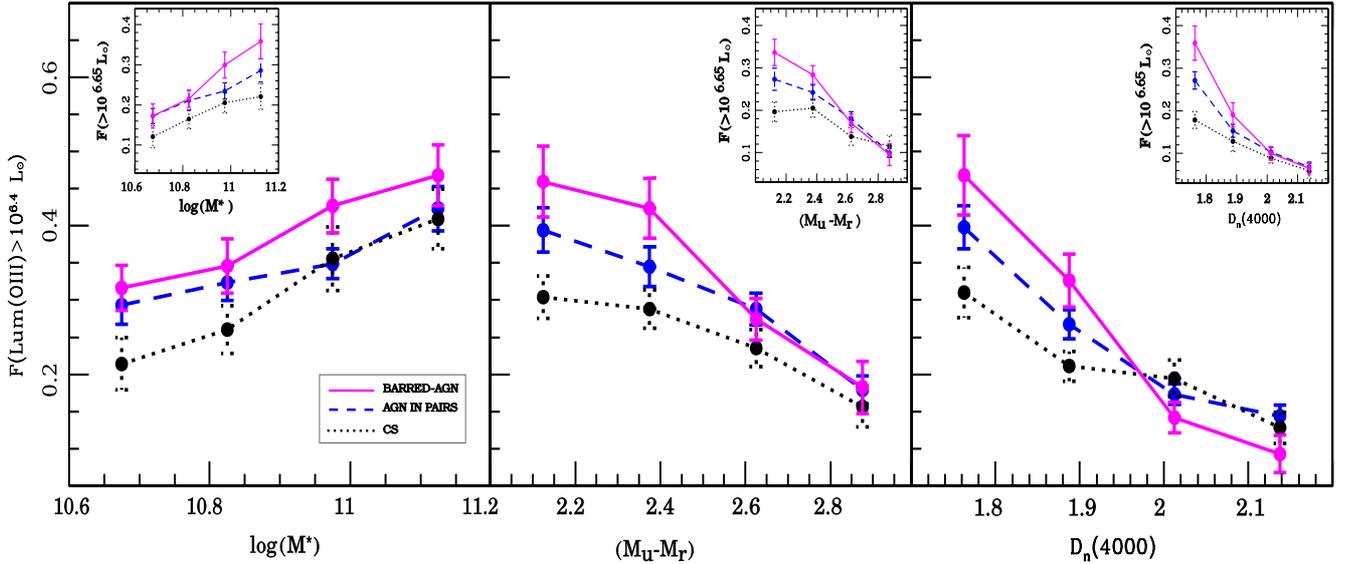}
\caption{Fraction of $Lum[OIII] > 10^{6.4} L_{\odot}$ as a function of stellar mass, ($M_u-M_r$) color, and $D_n(4000)$ parameter (left, middle and right panels, respectively), for barred AGN galaxies (solid lines), AGN in pair systems (dashed lines), and unbarred AGN in the control sample (dotted lines).
Each bin for each sample contains a similar number of galaxies.  
The number average of galaxies in each bin is $\approx$ 105 for \textsc{barred-agn} $\approx$  310 for \textsc{agn in pairs} and $\approx$ 140 for \textsc{CS}.
The small boxes correspond to the fraction of $Lum[OIII] > 10^{6.65} L_{\odot}$ as a function of the same parameters.
}
\label{LOprop}
\end{figure*}

In Fig.~\ref{LOprop}, we present the fraction of galaxies 
with $Lum[OIII] > 10^{6.4} L_{\odot}$ as a function of host galaxy properties, for barred AGN (solid lines), AGN in pair systems (dashed lines), and unbarred AGN galaxies in the control 
sample (dotted lines).
The left panel shows the fraction of galaxies with high $Lum[OIII]$ values as a function of stellar mass content of the corresponding host galaxies.
It can be clearly seen that, in general, the most massive hosts show a higher fraction of AGNs with $Lum[OIII] > 10^{6.4} L_{\odot}$. 
We can also observe that barred AGN objects show a slightly higher fraction of powerful AGN galaxies irrespective of the stellar mass range. Moreover, AGNs in interacting pair systems present a slight trend towards intermediate values between those for barred active galaxies and AGN hosts in the control sample.
This suggests a moderate enhancement of the nuclear activity for barred AGN with respect to AGN residing in pair systems although both samples exhibit an increment of high $Lum[OIII]$ values with respect to the unbarred hosts in the control sample,
regardless of the stellar mass content of the host galaxies. 
This result is consistent with the analogous study from OH et al. (2012) who found that AGN strength, given by the $[OIII]$ luminosity, is enhanced by the presence of a bar and linearly correlates with stellar mass.
The error bars in the figures were calculated using bootstrap error resampling \citep{barrow84}. This technique was performed with 1000 iterations for our analysis.

The middle panel of Fig.~\ref{LOprop} shows the fraction of AGN galaxies in the three samples analyzed in this paper, 
with $Lum[OIII] > 10^{6.4} L_{\odot}$, as a function of ($M_u-M_r$) color.  
From this figure, we can see a trend consistent with the increase of the powerful AGN galaxy fraction towards bluer AGN hosts, in the three samples studied in this work.
We can also see that the fraction of barred AGN galaxies with $Lum[OIII] > 10^{6.4} L_{\odot}$ is slightly higher than 
that of AGNs in pair systems and AGN hosts without bars in the control sample. With respect to the other two samples, active galaxies in pair systems present intermediate values of the fraction of galaxies with high $Lum[OIII]$ values.
These trends are more clear for objects with colors bluer than ($M_u-M_r$) $\approx$ 2.4.

A similar tendency is observed with respect to the stellar age parameter $(D_n(4000))$ (right panel of Fig. \ref{LOprop}) where 
the fraction of AGN galaxies with higher nuclear activity decreases toward older stellar populations. 
In addition, barred AGN galaxies present a steeper slope and a moderate increase of the powerful AGN galaxy fraction, with respect to the other AGN samples. This trend is more remarkable towards smaller $D_n(4000)$ values.
The fraction of $Lum[OIII] > 10^{6.4} L_{\odot}$ for the AGNs in interacting pairs is higher than that of the AGN in the control sample, although the former exhibit a slight tendency towards lower values with respect to the barred active galaxies, throughout the whole $D_n(4000)$ range.
In a similar way, A13 found that barred host galaxies systematically
show a higher fraction of powerful AGNs with respect to
the unbarred spiral objects of a suitable control sample, in bins of different galaxy properties.

In order to check these trends, we also calculated the fraction of galaxies with $Lum[OIII] > 10^{6.65} L_{\odot}$ as a function of the host galaxy properties, for the three samples studied (see small boxes in Fig. \ref{LOprop}).
We consider the limit $Lum[OIII] = 10^{6.65} L_{\odot}$ by selecting
the value that represents the excess of nuclear activity for \textsc{barred-agn} with respect to the control galaxies. 
This value also represents a significant difference between the percentages of both samples, maintaining a reliable number of objects in both catalogs.
This threshold is represented by the dotted vertical line in Fig. 3.
As can be observed, the fractions calculated with different limits maintain a similar behavior for the three samples (as a function of stellar mass, $M_u-M_r$ color, and $D_n(4000)$ parameter) implying stable results.

Based on numerical simulations \cite{robi17} study  the impact of AGN accretion and feedback on barred disk galaxies, showing that the feedback mainly affects the dynamics of the central region of the galaxies. 
Feedback pushes gas outward, colliding with the gas inflowing along the bar. As a result, early and efficient star formation is driven to larger radii.
In addition, unbarred compared to barred galaxies have a lower AGN luminosity, and besides, the central region contains less gas that could be affected by feedback.

\begin{figure}
\centering
\includegraphics[width=80mm,height=80mm ]
{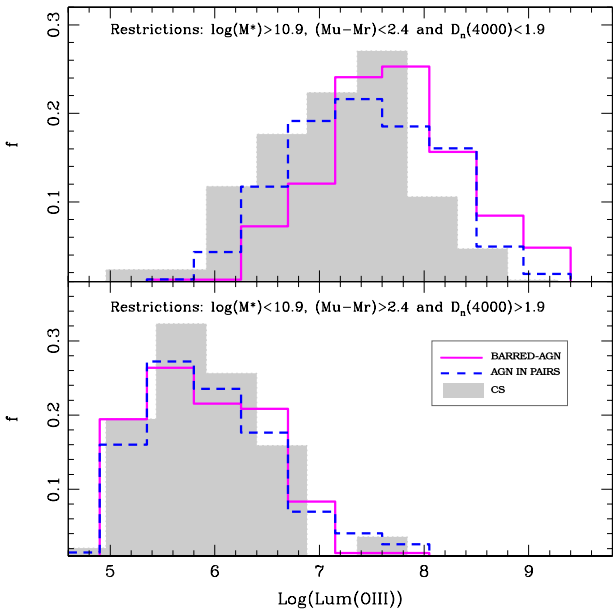}
\caption{Distributions of $log(Lum[OIII])$ for barred AGN galaxies (solid lines), AGNs in pair systems (dashed lines), 
and unbarred AGNs in the control sample (full surfaces), restricted in two ranges of stellar mass content, colors, and stellar age populations.
}
\label{HLOprop}
\end{figure}

Taking into account the results shown in Fig.6, we studied the distribution of $Lum[OIII]$ for a sample of AGN hosts with equal numbers of objects, considering low and high stellar mass content, blue/red colors, and young/old stellar populations.
The adopted thresholds were log($M^*$)=10.9, ($M_u-M_r$)=2.4, and $D_n(4000)$=1.9.
A similar threshold in stellar mass was used by Ellison et al. (2011a) finding that 
above  $M^* > 10^{10} M_{\odot}$,  the population of barred galaxies is dominated by early-type spirals, with a larger central bulge.
The analysis was performed for the three AGN samples considered in this work.  
From Fig. \ref{HLOprop}, it can easily be seen that AGNs with more massive hosts and with bluer and younger populations show a remarkable excess of efficient nuclear activity (top panel) with respect to active galaxies with low stellar mass, redder colors, and older stellar populations (bottom panel). 
We also found a significant excess of barred objects exhibiting higher values of $Lum[OIII]$, showing that barred AGN have a more efficient nuclear activity with respect to the other samples. 
This behavior is more noticeable in massive hosts that have an important blue and young stellar population, while in less massive objects with red colors and old stellar populations, the $Lum[OIII]$ distributions show similar trends in the different AGN samples analyzed.
Kauffmann et al. (2003) found that the weakest AGNs have stellar ages in the range of early-type galaxies, $D_n(4000) > $1.7.
More recently, \cite{lee12} showed that for AGN host galaxies with 2.0 $< (u - r) <$2.5, the nuclear activity is higher in barred galaxies than in non-barred ones, 
 implying that a bars can boost AGN activity when the host galaxies tend to be bluer. 
To quantify these results, we calculate the percentages of extremely powerful active galaxies for AGNs with massive hosts, and with bluer and younger populations, finding 83$\%$, 71$\%,$ and 52$\%$ for AGNs in different samples (\textsc{barred-agn}, \textsc{agn in pairs} and \textsc{CS}, respectively), while for active galaxies with low stellar mass, redder colors, and older populations, the three samples present $\approx$ 5$\%$ of the $Lum[OIII] > 10^{7.0} L_{\odot}$.

\begin{figure*}
\centering
\includegraphics[width=180mm,height=60mm ]{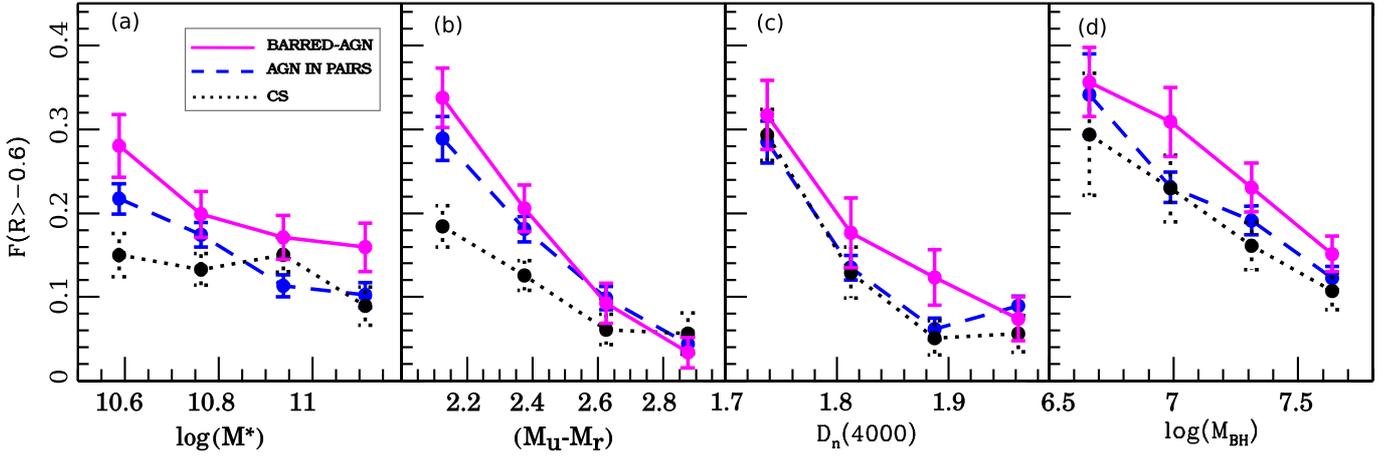}
\caption{Fraction of $R > -0.6$ as a function of stellar mass, ($M_u-M_r$) color, $D_n(4000)$ parameter,or $log(M_{BH})$ 
($a$, $b$, $c$, and $d$, respectively), for barred AGN galaxies (solid lines), AGNs in pair systems (dashed lines), and unbarred AGNs in the control sample (dotted lines).
Each bin for each sample contains a similar number of galaxies. The average number of galaxies in each bin is $\approx$ 93 for \textsc{barred-agn}, $\approx$  270 for \textsc{agn in pairs} and $\approx$ 110 for \textsc{CS}.
}
\label{Rprop}
\end{figure*}

To complete the analysis, in Fig. \ref{Rprop} we show the fraction of galaxies with $\cal R$$>-0.6$ as a function 
of stellar mass, $(M_u-M_r)$ color, stellar age indicator, or  black hole mass, 
for AGNs with bars, AGNs in pair systems and AGNs in the control sample (solid, dashed, and dotted lines, respectively).
In this analysis, for the three samples, we used the $\cal R$ parameter calculated with the same values of $\alpha$ and $\beta$
as we used for the unbarred AGN hosts,  
in order to make a real and effective comparison between the AGN galaxies in the different catalogs.
The fraction of AGN galaxies with  $\cal R$$>-0.6$ decreases with stellar mass (panel $a$), $(M_u-M_r)$ color (panel $b$), and 
stellar population age (panel $c$), for the three samples. 
Barred AGN galaxies can be seen to show a slight tendency towards higher accretion rates than the other AGN samples. Also, AGNs belonging to pair systems exhibit an increase in objects with $\cal R$$>-0.6$ with respect to active galaxies in the control sample. 
This result suggests that more massive and bluer AGN host 
galaxies with a younger stellar population  have significantly higher accretion rates than less massive and redderAGN hosts with older stellar populations, meaning that that gas-rich galaxies are more efficient in accreting gas toward the central black hole. 
In this context, bar perturbations and interactions between galaxies play an important role. 

On the other hand, \cite{lee12} measured the median values of log($Lum[OIII]$/$M_{BH})$ as a function of $(u-r)$, and as a function of $log(M^*),$ for barred and unbarred AGNs selected using the \cite{kel01} classification, also from the SDSS-DR7 volume-limited sample of late-type galaxies (b/a $>$ 0.6). 
These authors showed that median curves for strongly barred and weakly barred galaxies lie slightly above those for non-barred galaxies.
\cite {gal15} also studied the relative accretion strengths, $\cal R$, as a function of mass and color for barred and unbarred AGN galaxies, showing similar $\cal R$ values for both samples and an inverse relation of the acretion rate with the stellar masses and $(u-r)$ colors of the hosts, concluding that there is no strong evidence for a difference in accretion strength between barred and unbarred AGNs.
Additionally, \cite{gal15} exclude composites and LINERs from their
sample of AGNs.
The discrepancy in the results could arise mainly from differences in the AGN selection criteria, that may affect the results related to the effect of enhanced AGN activity due to bars.

As we can observe in panel (d) of Fig.\ref{Rprop}, in general, objects with   smaller black holes are those that exhibit the higher fraction of $\cal R$$>-0.6$. 
Since stellar mass is strongly correlated with black hole mass (H{\"a}ring \& Rix 2004; G{\"u}ltekin et al. 2009; Merloni et al. 2010), this explains the trends seen in both parameters for our samples.
Furthermore, AGNs in barred host galaxies show a moderate excess of accretion rate in comparison with the other samples in the whole $log(M_{BH})$ range, while AGNs in pairs present slightly higher values of $\cal R$ with respect to the AGN galaxies in the control sample.

AGNs in barred hosts and AGNs belonging to a galaxy in a pair show      an excess of high nuclear activity and higher accretion rates with respect to AGNs in the control sample. These findings imply that bars and interactions can aid the gas infall onto the central black holes. Moreover, the fact that AGNs in barred galaxies exhibit an excess of $\cal R$ and $Lum[OIII]$ with respect to  AGNs in pair systems shows that the bar perturbations may be a more efficient mechanism to transport material towards the inner central regions than the radial instabilities induced by galaxy interactions.

The different timescales of these phenomena should also be highlighted: the short phase of interactions compared to the longer life spans of bars (Ellison et al. 2011a).   
In this context, our results may suggest that the internal mechanism (bar perturbation in the disk) leads to more effective transport of the radial gas flow onto the innermost zones of the galaxies than galaxy interactions. 
The longer life span of bars could be crucial in producing gas inflows for a long period, which could maintain  an efficient central nuclear activity.   

%------------------------------------------------------------------------------------
\section{The role of a companion galaxy}

We have also investigated the role played by a pair companion in powering the nuclear activity.
This was accomplished by analyzing the OIII luminosity, $Lum[OIII],$ of AGNs in pair systems as a function of the properties of the galaxy pair companion.
For the following analysis, we extracted the galaxy companion of the AGNs in the \textsc{agn in pairs} sample described in Sect. 2.2.
We reiterate that  $\sim$ 20$\%$ of this sample corresponds to systems composed of two AGN.    
In this context, Ellison et al. (2011b) studied the fraction of paired systems with two AGNs using a statistical approach, finding a larger fraction at small projected separations, remaining high out to 80-170 kpc $h^{-1}$.
More recently, Fu et al. (2018) analyzed close pairs with projected separations between 1 and 30 kpc $h^{-1}$, and radial velocity differences less than 600 km $s^{-1}$, finding that the fraction of binary AGNs is  $\sim$ 13$\%$ of the total pair sample.
We will perform a study of AGN-AGN pairs in future work.

\begin{figure}
\centering
\includegraphics[width=80mm,height=100mm ]{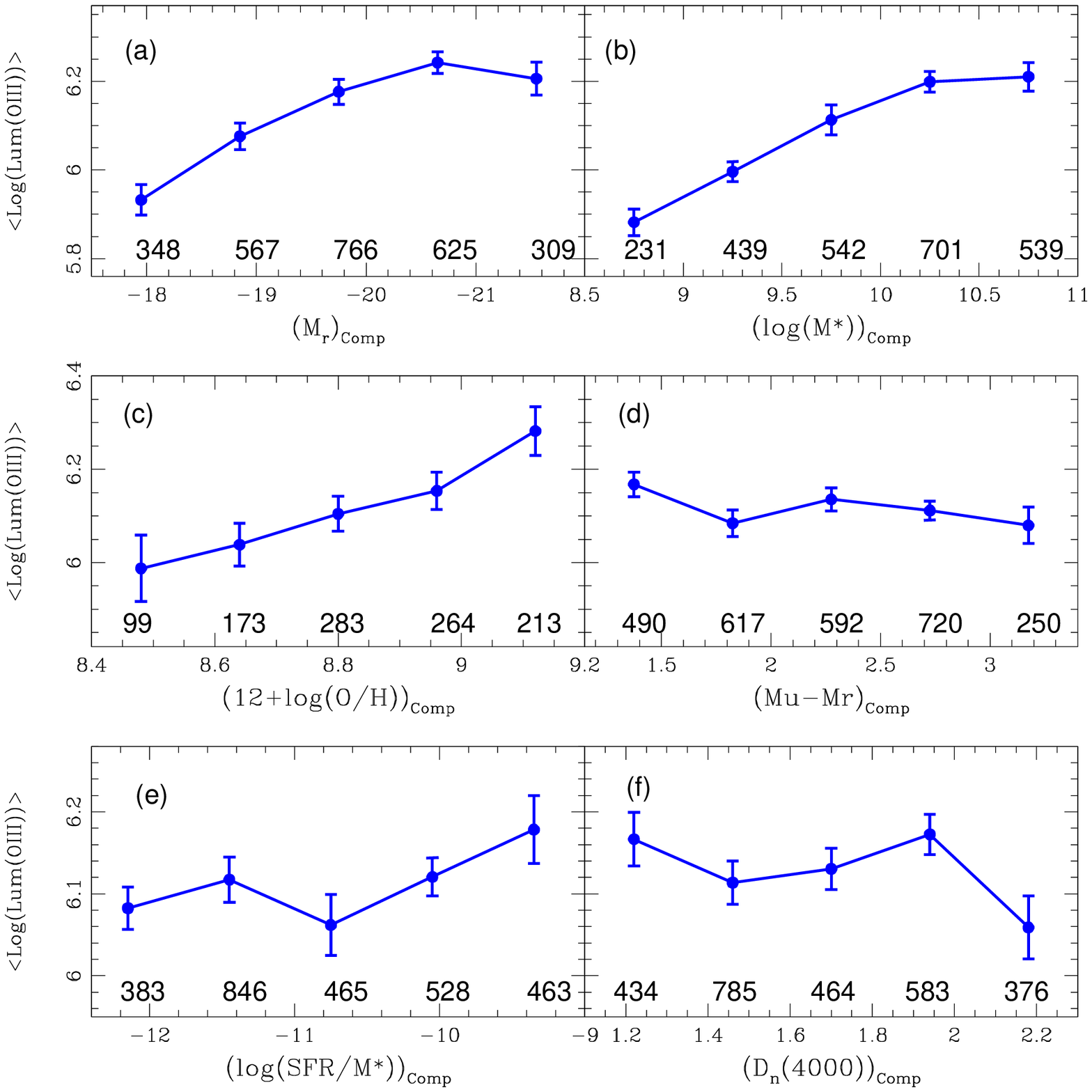}
\caption{$< log(Lum[OIII]) >$ of AGN galaxies in pair systems as a function of $M_r$,  $log(M^*)$, $12+log(O/H)$, $(M_u-M_r)$,
$log(SFR/M*)$ or $D_n(4000)$ of the galaxy pair companion. 
The number of galaxies in each bin is inset at the bottom of the figure.
}
\label{OIIIcomp}
\end{figure}

The results shown in Fig. \ref{OIIIcomp} indicate
that AGN activity depends significantly on the properties of the galaxy pair companion. 
Nuclear activity in galaxies with bright and more massive companions is significantly enhanced with respect to those with faint and less massive companions (see panels (a) and (b)). 
Even though galaxy mass and luminosity are directly correlated, the spread in the relation suggests that it would be beneficial to study both parameters separately to search for possible differences. We find the relation between the magnitude of the galaxy companion and the AGN activity is consistent with the mass dependence, as expected.

In line with this, \cite{alo07} studied active galaxies in close pair systems, considering  pairs with bright and faint companions separately, adopting $M_r =-20$ as the magnitude threshold. They also found that AGN activity in hosts with bright companions show a clear increase in $Lum[OIII]$ with respect to the nuclear activity in AGNs with faint companions. 
We also analyze the metallicity of the galaxy companion and its effect on nuclear activity. As a metallicity parameter, we used $12+log\left(O/H\right)$ which represents the ratio between oxygen and hydrogen abundances (Tremonti et al. 2004). This parameter principally reflects the amount of gas reprocessed by the stars and depends strongly on the evolutionary state of a galaxy. 
As is expected from the well-known mass-metallicity relation \citep{tremonti04,kel08}, it can be observed that galaxies with higher metallicities have a companion with stronger nuclear activity (see panel (c)). This result complements the studies from \cite{alo07}, in the sense that major interactions induce stronger AGN activity.

In addition, we analyze colors, star formation activity, and stellar age population of the AGN  companions in galaxy pairs, with the aim to study the relation between these properties and the nuclear activity of the neighboring AGN. 
In the following analysis we use the specific SFR parameter, $log\left(SFR/M_{*}\right)$ as a good indicator of the star formation activity; it is estimated as a function of the $H\alpha$ line luminosity, and normalized using stellar mass (Brinchmann et al. 2004).  
We also plot in  Fig. \ref{OIIIcomp} the $<log(Lum[OIII])>$ of the AGNs in pairs as a function of $(M_u-M_r)$, $log(SFR/M_{*})$ and $D_n(4000)$ of the galaxy pair companions (panels (d), (e) and (f)). This figure shows that AGNs in pairs present an increase in the OIII luminosity when the galaxy companions exhibit blue colors\textbf{\footnote{ The $\Delta$$\chi^{2}$ test between a constant and a linear fit to the color data indicates that the $Lum[OIII]$ is consistent with a tendency toward bluer colours ($\Delta$$\chi^{2}$ $=$ 14.04).}}, 
efficient star formation activity, and a young stellar population.
This finding provides evidence that nuclear activity in AGN objects is affected by the presence of the close companion, and the properties of this near galaxy are also an important issue to take into account.

\begin{figure*}
\centering
\includegraphics[width=180mm,height=105mm ]{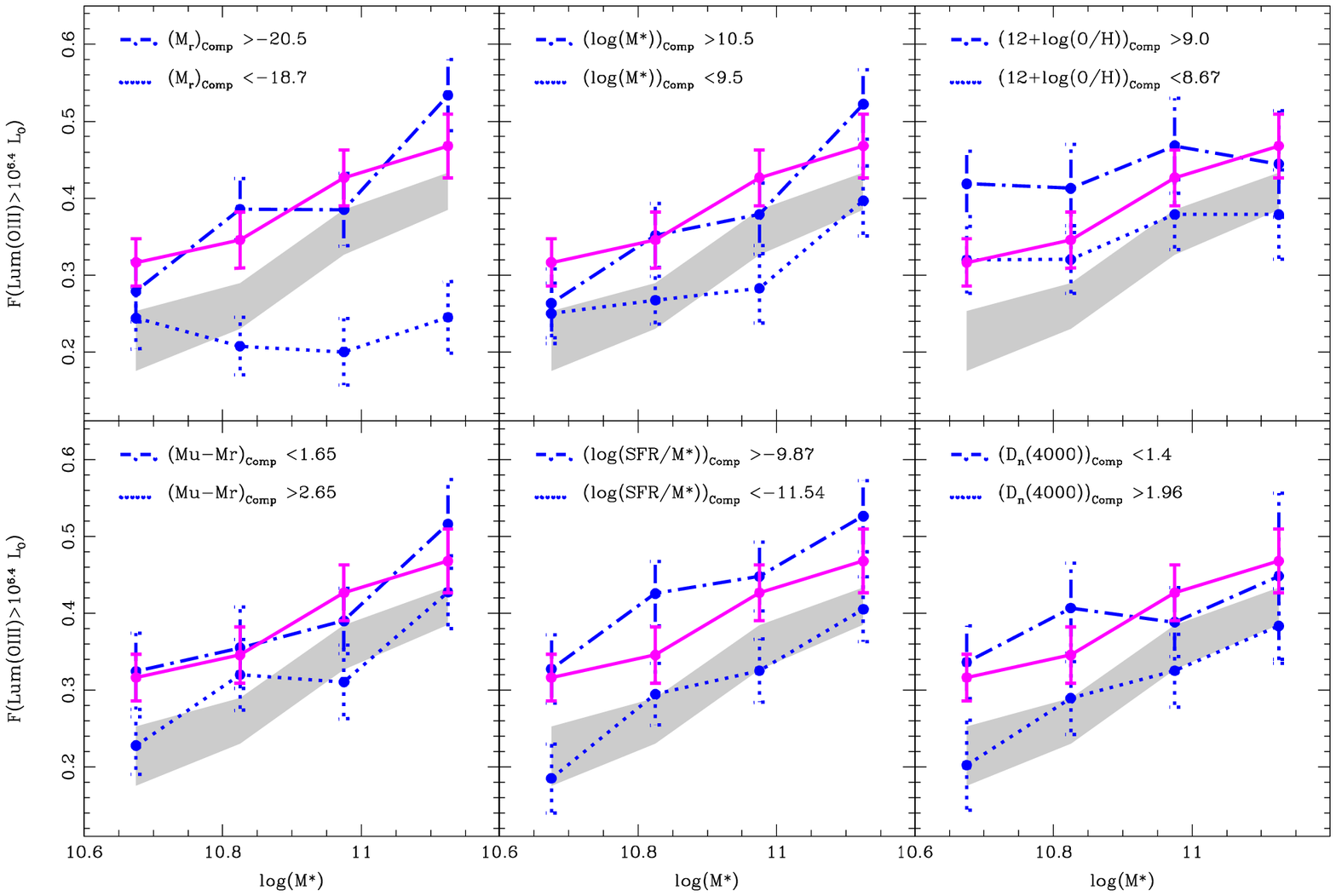}
\caption{Fraction of $Lum[OIII] > 10^{6.4} L_{\odot}$ as a function of stellar mass, $log(M^*)$, for barred AGN galaxies 
(solid lines) and AGNs in pairs with restrictions in the galaxy pair companion properties defined in the text 
(dotted and dot-dashed lines).  
The gray surfaces represent the control sample within uncertainties derived
through the bootstrap resampling technique.
Each bin for each sample contains a similar number of galaxies.  
The average number of galaxies in each bin is $\approx$ 105 for \textsc{barred-agn}, $\approx$ 140 for \textsc{CS} and $\approx$ 67 for \textsc{agn in pairs} with restrictions in the properties of the galaxy pair companion.
}
\label{OIIIpropC}
\end{figure*}

In Fig. \ref{OIIIpropC} we plot the fraction of AGN galaxies in pair systems with high OIII luminosity 
($Lum[OIII] > 10^{6.4} L_{\odot}$) as a function of their stellar 
mass content. We consider differences in the host properties of galaxy companions by analyzing the first ($Q_1$) and fourth ($Q_4$) quartile of the distributions of absolute magnitude, $M_r$, stellar mass, $log(M^*)$, metallicity, $12+log(O/H)$, $(M_u-M_r)$ color,  $log(SFR/M_{*})$, and  $D_n(4000)$ of the galaxy pair companions  of the AGNs (see Table 3). 
We also display the $Lum[OIII] > 10^{6.4} L_{\odot}$ for barred active galaxies and AGNs in the control sample.  
The results shown in this figure clearly indicate that nuclear activity is significantly enhanced in AGN hosts with galaxy companions that are brighter and
more massive, and have  higher metallicity, bluer colors, more efficient star formation and a younger stellar population, 
which show a similar fraction of $Lum[OIII]$ values than those of barred active galaxies.
Inversely, AGNs with galaxy companions that present low masses, luminosities, and metallicities, red colors, an old stellar population 
and less efficiency in forming stars display low nuclear activity. In addition, these AGN galaxies show similar (lower in some cases) 
$Lum[OIII] > 10^{6.4} L_{\odot}$ values, as a function of the $log(M^*)$, to active galaxies in the control sample.     
Moreover, in Table 3 we quantify the percentage of the AGNs in pair systems with high- and extreme- nuclear activity ($Lum[OIII] = 10^{6.4} L_{\odot}$ and $Lum[OIII] = 10^{7} L_{\odot}$), taking into account the galaxy pair companion properties limited for the first and fourth quartiles of the $M_r$, $log(M^*)$, $12+log(O/H)$, $(M_u-M_r)$, $log(SFR/M_{*})$, and $D_n(4000)$ distributions, following the previous analysis. The values of the percentages reflect the findings shown in Fig. \ref{OIIIpropC}.

These results clearly indicate that the efficiency of the mergers and interactions in transporting material towards the inner 
regions of the galaxies depends not only on the properties of the hosts, but is also strongly influenced by the 
galaxy pair companion characteristics. 
In addition, when the galaxy companion tends to be massive, luminous, and with high gas content (represented by the host properties of the neighbor galaxy), the effect produced by the external mechanism of mergers and interactions on the central nuclear activity tends to be as efficient as that induced by bars.

\begin{table*}
\center
\caption{Percentages of AGN galaxies in pair systems with high luminosity and extremely powerful nuclear activity, taking into account the properties of the galaxy pair companion.
}
\begin{tabular}{|c c c | }
\hline
Restrictions & $L[OIII] > 10^{6.4} L_{\odot}$ & $L[OIII] > 10^{7.0} L_{\odot}$ \\
\hline
\hline
$Q_1$: ($M_r$)$_{Comp}$ $<$ -20.5     &   58.4$\%$ $\pm$ 1.1 &   34.8$\%$ $\pm$ 2.2  \\
$Q_4$: ($M_r$)$_{Comp}$ $>$ -18.7     &   42.7$\%$ $\pm$ 0.9 &   20.4$\%$ $\pm$ 1.9  \\
\hline
$Q_1$: ($log(M^*)$)$_{Comp}$ $<$ 9.5  &    47.5$\%$ $\pm$ 1.0 &   25.3$\%$ $\pm$ 1.6  \\
$Q_4$: ($log(M^*)$)$_{Comp}$ $>$ 10.5  &   55.8$\%$ $\pm$ 1.2 &   33.5$\%$ $\pm$ 1.7  \\
\hline
$Q_1$: (12+log(O/H)$)_{Comp}$ $<$ 8.67 &  51.0$\%$ $\pm$ 1.2 &  28.4$\%$ $\pm$ 2.6  \\
$Q_4$: (12+log(O/H)$)_{Comp}$ $>$ 9.0 &   62.2$\%$ $\pm$ 1.3 &  38.7$\%$ $\pm$ 2.5  \\
\hline
$Q_1$: ($(M_u-M_r)$)$_{Comp}$ $<$ 1.65 &  55.9$\%$ $\pm$ 0.8 &   32.7$\%$ $\pm$ 2.7  \\
$Q_4$: ($(M_u-M_r)$)$_{Comp}$ $>$ 2.65 &  49.5$\%$ $\pm$ 0.9 &   27.9$\%$ $\pm$ 2.5  \\
\hline
$Q_1$: ($log(SFR/M^*)$)$_{Comp}$ $<$ -11.54 & 50.7$\%$ $\pm$ 1.4 &  27.8$\%$ $\pm$ 2.4  \\
$Q_4$: ($log(SFR/M^*)$)$_{Comp}$ $>$ -9.87 &  58.6$\%$ $\pm$ 1.2 &  35.2$\%$ $\pm$ 2.4  \\
\hline
$Q_1$: ($D_n(4000)$)$_{Comp}$ $<$ 1.4 &   54.5$\%$ $\pm$ 1.0 &  34.5$\%$ $\pm$ 1.8  \\
$Q_4$: ($D_n(4000)$)$_{Comp}$ $>$ 1.96 &  51.0$\%$ $\pm$ 0.9 &  28.1$\%$ $\pm$ 2.1  \\
\hline
\end{tabular}

{\small Note: $Q_1$ and $Q_4$ correspond to the first and the fourth quartile of the distributions of the pair companion properties.}
\end{table*}

%-----------------------------------------------------------------------------

\section{Summary and Conclusions}

We have performed a comparative analysis of the effect of bars and mergers/interactions on the central nuclear activity of spiral AGN galaxies. 

We acknowledge that different AGN selection criteria can lead to the selection of galaxies with  different SFR activity. Low-luminosity radio-selected
AGN are significantly biased towards low SFR values compared 
to IR selected galaxies in the SDSS. Ellison et al. (2016) suggest the dominance of mergers in IR selected AGNs, a lower merger incidence amongst optically selected AGNs, and that secular fuelling dominates low-excitation radio galaxies (LERGs).

Our study is based on homogeneous samples of optically selected AGNs with strong bars, in relative isolation or in pair systems.
In order to carry out a suitable comparison of the effects of bars and mergers/interactions, we selected AGN spiral galaxies in both samples with similar redshift, $g-$band apparent magnitude, absolute $r-$band magnitude, stellar mass, color, and stellar age population distributions. 
To obtain an appropriate quantification of the effect of the two processes (bars and interactions) on the nuclear activity, we also constructed a suitable control sample of unbarred spiral AGN galaxies, without a pair companion, with similar host properties to those of the other samples.

The main results and conclusions of our analysis are summarized as follows.

We found that barred active galaxies show an excess of nuclear activity compared to AGN galaxies in pair systems. 
Moreover, both samples show an excess of high $Lum[OIII]$ values with respect to unbarred spiral active galaxies in the control sample.

We also analyzed the accretion strength onto a central black hole for AGN host galaxies in the different samples. From this study, we conclude that barred active galaxies have an excess of objects with high accretion rate values with respect to AGN hosts inhabiting pair systems.
Furthermore, active galaxies in both samples exhibit higher $\cal R$ values than the control active galaxies.  

We studied the fraction of powerful AGN galaxies ($Lum[OIII] > 10^{6.4} L_{\odot}$) as a function of stellar mass, color, and stellar age population.
We found that the number of active galaxies with efficient central nuclear activity increases when selecting host galaxies with larger stellar mass, bluer colors, and younger stellar populations.
From this analysis, we also show that the fraction of $Lum[OIII] > 10^{6.4} L_{\odot}$ for the AGNs in pairs is slightly higher than that of the AGNs in the control sample, throughout the whole range of  $log(M^*)$, $(M_u-M_r)$ and $D_n(4000)$.
Concurrently, we found that barred active objects show a moderately higher fraction of powerful AGNs with respect to the other AGN samples, in bins of different galaxy properties.  

We also explored the fraction of AGN galaxies with high accretion rate ($\cal R$$>-0.6$) as a function of the host galaxy properties.
We found that the fraction of active galaxies with  $\cal R$$>-0.6$ increases towards more massive hosts with bluer colors and younger stellar populations. 
From this analysis we also show that barred AGN host objects exhibit a slightly higher fraction of efficient accretion rate with respect to the other samples, in bins of different galaxy properties.
We also found that AGN inhabiting pair systems display higher $\cal R$$>-0.6$ values in comparison with active galaxies in the control sample.
Furthermore, we found that the smallest black holes exhibit a higher 
fraction of $\cal R$$>-0.6$, and, in this context, barred AGN galaxies show a moderate excess of efficient accretion rate with respect to active galaxies in the other samples, throughout the whole range of the  $log(M_{BH})$. 

For AGNs inhabiting pair systems, we found that the nuclear activity is remarkably dependent on the galaxy pair companion properties.
In this context, we show that the nuclear activity of an AGN companion presents a noticeable increment when its galaxy host is brighter and more massive, with higher metallicity, bluer colors, more efficient star formation activity and a young stellar population.

We also explored the fraction of AGNs in pair systems with higher OIII luminosity as a function of $log(M^*)$, considering galaxy companions with different host features separately.  
We found that the fraction of $Lum[OIII] > 10^{6.4} L_{\odot}$, for AGNs belonging to pair systems, is similar to that of barred active galaxies when galaxy pair companions present bright and massive hosts with high metallicity, blue colors, efficient star formation activity and a young stellar population.
Conversely, the fraction of powerful nuclear activity is similar for AGNs in pairs and for AGNs in the control sample, when the galaxy pair companions present low masses, luminosities and metallicities, red colors,  less star formation activity and an old stellar population.  

The results found in this work suggest an important effect of both  bars and interactions in driving radial gas inflows towards the innermost regions of galaxies. These mechanisms produce an enhancement in nuclear activity and accretion rate of central black holes in spiral active nuclei galaxies. 
It should also be noted that the internal process of the bar perturbation  presents a more effective transport of the gas flow to the central zones in comparison with the external mechanism of the mergers and interactions. 
Furthermore, the impact of the AGNs belonging to pair systems on the central nuclear activity is noticeably influenced by the galaxy pair companion properties.

\begin{acknowledgements}
Authors would like to thank an anonymous referee who contributed insightful suggestions. 

This work was partially supported by the Consejo Nacional de Investigaciones
Cient\'{\i}ficas y T\'ecnicas and the Secretar\'{\i}a de Ciencia y T\'ecnica 
de la Universidad Nacional de San Juan.

Funding for the SDSS has been provided by the Alfred P. Sloan
Foundation, the Participating Institutions, the National Science Foundation,
the U.S. Department of Energy, the National Aeronautics and Space
Administration, the Japanese Monbukagakusho, the Max Planck Society, and the
Higher Education Funding Council for England. The SDSS Web Site is
http://www.sdss.org/.

The SDSS is managed by the Astrophysical Research Consortium for the
Participating Institutions. The Participating Institutions are the American
Museum of Natural History, Astrophysical Institute Potsdam, University of
Basel, University of Cambridge, Case Western Reserve University,
University of
Chicago, Drexel University, Fermilab, the Institute for Advanced Study, the
Japan Participation Group, Johns Hopkins University, the Joint Institute for
Nuclear Astrophysics, the Kavli Institute for Particle Astrophysics and
Cosmology, the Korean Scientist Group, the Chinese Academy of Sciences
(LAMOST), Los Alamos National Laboratory, the Max-Planck-Institute for
Astronomy (MPIA), the Max-Planck-Institute for Astrophysics (MPA), New Mexico
State University, Ohio State University, University of Pittsburgh, University
of Portsmouth, Princeton University, the United States Naval Observatory, and
the University of Washington.
\end{acknowledgements}


\begin{thebibliography}{abbrvnat.bst}
%\begin{thebibliography}{}

\bibitem[Abazajian \etal (2009)]{aba09} Abazajian K. N.; Adelman-McCarthy J. K.; Agüeros M. A.; Allam S. S.; Allende Prieto C.; An D.; Anderson K. S. J.; Anderson S. F., et al. 2009, ApJS, 182, 543.  

\bibitem[Alonso et al. (2006)]{alo06} Alonso M.S., Lambas D.G., Tissera P.B. \& Coldwell G., 2006, \mnras, 367, 1029.

\bibitem[Alonso \etal (2007)]{alo07} Alonso M.S., Lambas D.G., Tissera P. and Coldwell G., 2007, MNRAS 375, 1017. 

\bibitem[Alonso et al. (2012)]{alo12} Alonso S., Mesa V., Padilla N., Lambas D. G., 2012, \aap, 539A, 46A.

\bibitem[Alonso \etal (2013)]{alo13} Alonso S., Coldwell G. , Lambas D. G., 2013, A\&A, 549, 141. 

\bibitem[Alonso \etal (2014)]{alo14} Alonso S., Coldwell G. , Lambas D. G., 2014, A\&A, 572, 86.   

\bibitem[Athanassoula (1983)]{atha83} Athanassoula, E. 1983, IAU Symp., 100, 243.

\bibitem[Baldwin, Phillips \& Terlevich (BPT, 1981)]{BPT81} Baldwin J. A., Phillips M. M. \& Terlevich R., 1981, \pasp, 93, 5.

\bibitem[Balogh \etal (1999)]{balo99} Balogh M., Morris, S. L., Yee, H. K. C., Carlberg, R. G., Ellingson, E., 1999, \apj, 527, 54.   

\bibitem[Balogh \etal (2004)]{balo04} Balogh M., Baldry I. K., Nichol R., Miller C., Bower R., 
Glazebrook K., 2004, \apj Letters, 615, 101.

\bibitem[Barnes \& Herquist (1996)]{bahe96} Barnes J. E., Herquist L., 1996, ApJ, 471, 115. 

\bibitem[Barrow, Bhavsar \& Sonoda (1984)]{barrow84}Barrow J.D., Bhavsar S.P. \& Sonoda B.H.,
1984, \mn, 210, 19. 

\bibitem[Barrows et al.(2017)]{bar17} Barrows, R.~S., Comerford, J.~M., Zakamska, N.~L., \& Cooper, M.~C.\ 2017, \apj, 850, 27.

\bibitem[Barton, Geller \& Kenyon  (2000)]{barton} Barton E.J., Geller M.J. \& Kenyon S.J., 2000, \apj, 530, 660.

\bibitem[Blanton \etal (2005)]{blanton05}Blanton M.R., Eisenstein D., Hogg D.W., Schlegel D.J  \& Brinkmann J., 2005 \apj, 629, 143.

\bibitem[Benn \etal (2008)]{ben08} Bennert N., Canalizo G., Jungwiert B., Stockton A., Schweizer F., Peng  C. Y., Lacy M., 2008, ApJ, 677, 846.

\bibitem[Best \& Heckman (2012)]{byh12} Best P.N. \& Heckman P.M., 2012, \mn, 421, 1569. 

\bibitem[Brinchmann \etal (2004)]{brinch04} Brinchmann J., Charlot S., White S. D. M., Tremonti C., Kauffmann G., Heckman T. \& Brinkmann J., 2004, \mn, 351, 1151. 

\bibitem[Buta \& Combes (1996)]{buta96} Buta, R., \& Combes, F. 1996, Fund. Cosm. Phys., 17, 95.

\bibitem[Calzetti \etal (2000)]{calzetti00} Calzetti D., Armus L., Bohlin R.C.,
Kinney A.L., Koornneef J. \& Storchi-Bergmann T., 2000, \apj, 533, 682.

\bibitem[ Canalizo \& Stockton (2001)]{can01}  Canalizo G., Stockton A., 2001, ApJ, 555, 719. 

\bibitem[Carollo \etal (2002)]{caro02} Carollo C. M., Stiavelli M., Seigar M., de Zeeuw P. T., Dejonghe H., 2002, AJ, 123, 159.  

\bibitem[Cheung \etal (2013)]{cheu13} Cheung E., Athanassoula E., Masters K., Nichol R., Bosma A., Bell E.F., Faber S., Koo D., \etal, 2013, AJ, 779, 162.

\bibitem[Cheung et al.(2015)]{cheu15} Cheung, E., Trump, J.~R., Athanassoula, E., et al.\ 2015, \mnras, 447, 506.

\bibitem[Combes \etal (1993)]{comb93} Combes F., Elmegreen B. G., 1993, A\&A, 271, 391.  

\bibitem[Corsini \etal (2003)]{cor03} Corsini E. M., Debattista V. P., Aguerri J. A. L., 2003, ApJ, 599, 29.  

\bibitem[Coldwell \etal (2009)]{coldwell09} Coldwell G.V., Lambas D.G., S{\"o}chting I.K. \& Gurovich S., 2009, \mn, 399, 88.

\bibitem[Coldwell et al.(2014)]{coldwell14} Coldwell, G.~V., Gurovich, S., D{\'{\i}}az Tello, J., S{\"o}chting, I.~K., \& Lambas, D.~G.\ 2014, \mnras, 437, 1199.

\bibitem[Di Matteo, Springel \& Hernquist (2005)]{dimat05} Di Matteo T., Springel V. \& Hernquist L., 2005, Nature, 433, 604.

\bibitem[Du et al.(2017)]{du17} Du, M., Debattista, V.~P., Shen, J., Ho, 
L.~C., \& Erwin, P.\ 2017, \apjl, 844, L15.

\bibitem[Ellison et al. (2011a)]{ell11a} Ellison S. L., Patton D. R., Mendel J. T., Scudder J. M., 2011, MNRAS, 418, 2043.  

\bibitem[Ellison et al. (2011b)]{ell11b} Ellison S., Nair P., Patton D., Scudder J.M., Mendel J.T., Simard L., 2011, MNRAS, 416, 2182.

\bibitem[Ellison et al. (2016)]{ell16} Ellison S., Hossen T., Rosario D. J., Mendel J. T.,  2016, MNRAS, 458, 34.

\bibitem[Ellison et al. (2013)]{ell13} Ellison S. L., Mendel J. T., Scudder J. M., Patton D. R., Palmer M. J. D., 2013, MNRAS, 430, 3128. 

\bibitem[Emsellem \etal (2001)]{emse01}Emsellem E., Greusard D., Combes F., Friedli D., Leon S., P\'econtal E. \& Wozniak H., 2001, \aap, 368, 52.

\bibitem[Fu et al.(2018)]{fu18} Fu H., Steffen J. L., Gross A. C., Dai Y.S., Isabell J. W., Lin L., Wake D., Xue R., et al. 2018, ApJ, 856, 93.

\bibitem[Galloway et al.(2015)]{gal15} Galloway, M.~A., Willett, K.~W., Fortson, L.~F., et al.\ 2015, \mnras, 448, 3442.

\bibitem[Goulding et al.(2017)]{gou17} Goulding, A.~D., Matthaey, E., Greene, J.~E., et al.\ 2017, \apj, 843, 135.

\bibitem[Graham (2008)]{gra08} Graham A. W., 2008, \aj, 680, 143.

\bibitem[G{\"u}ltekin \etal (2009)]{gulte09} G{\"u}ltekin, K., \etal, 2009, \apj, 706, 404.

\bibitem[[H\"aring \& Rix (2004)]{hari04} H\"aring N. \& Rix H.W., 2004, ApJ, 604, L89.

\bibitem[Heckman \etal (2004)]{heck04} Heckman T. M., Kauffmann G., Brinchmann J., Charlot S., Tremonti C., White S. D. M., 2004, \apj, 613, 109.

\bibitem[Heckman \etal (2005)]{heck05} Heckman T. M., Ptak A., Hornschemeier A., Kauffmann G., 2005, \apj, 634, 161.

\bibitem[Hickox, R. C., et al. (2009)]{hic09} Hickox R. C., et al., 2009, ApJ, 696, 891.

\bibitem[Jung et al.(2018)]{ju18} Jung, M., Illenseer, T.~F., \& Duschl, W.~J.\ 2018, arXiv:1802.06873).

\bibitem[Kauffmann \etal (2003)]{kauff03} Kauffmann G., Heckman T. M., Tremonti C. \etal, 2003, \mn, 346, 1055.

\bibitem[Kazantzidis \etal (2008)]{kaza08} Kazantzidis, S., Bullock, J. S., Zentner, A. R., Kravtsov, A. V. \& Moustakas, L. A. 2008, ApJ, 688, 254.

\bibitem[Kennicutt (1998)]{kenni} Kennicutt R., 1998, ARA \&A,  36, 189

\bibitem[Kewley \etal (2001)]{kel01} Kewley L. J., Dopita M. A., Sutherland R. S., et al. 2001, ApJ, 556, 121.

\bibitem[Kewley \& Ellison (2008)]{kel08} Kewley L.J. \& Ellison S.L., 2008, \apj, 681, 2.

\bibitem[Koss \etal (2010)]{koss} Koss M., Mushotzky R., Veilleux S., Winter L., 2010, ApJ, 716, L125.  

\bibitem[Koss \etal (2012)]{koss12} Koss M., Mushotzky R., Treister E., Veilleux S., Vasudevan R., Trippe M., 2012, ApJ, 746, L22.

\bibitem[Laine \etal (2002)]{laine02} Laine S., Shlosman I., Knapen J. H., Peletier R. F., 2002, \apj, 567, 97. 

\bibitem[Lambas et al. (2003)]{Lam03} Lambas D.G., Tissera P.B., Alonso M.S. \& Coldwell G., 2003, \mnras 346, 1189.

\bibitem[Lambas \etal (2012)]{Lam12} Lambas D. G., Alonso S., Mesa V., O'Mill A. L., 2012,  A\&A, 539, 45.  

\bibitem[Lee \etal (2012)] {lee12} Lee G., Woo J., Lee M.G., Hwang H.S., Lee J.C., Sohn J. \& Lee J.H.,2012, \apj, 750, 141.

\bibitem[Lynden-Bell (1969)]{LB69} Lynden-Bell D., 1969, \nat, 223, 690. 

\bibitem[Lintott et al.(2008)]{zoo} Lintott C.~J., Schawinski K., Slosar A., et al.\ 2008, \mnras, 389, 1179.

\bibitem[Lintott et al.(2011)]{zoo2} Lintott C., Schawinski K., Bamford S., et al.\ 2011, \mnras, 410, 166. 

\bibitem[Maciejewski \& Sparke(2000)]{mac2000} Maciejewski, W., \& Sparke, L.~S.\ 2000, \mnras, 313, 745. 

\bibitem[Malkan \etal (1998)]{malk98} Malkan M. A., Gorjian V., Tam R., 1998, ApJS, 117, 25.  

\bibitem[Martin (1995)]{martin95} Martin P., 1995, \aj, 109, 2428.

\bibitem[Martin et al.(2018)]{mar18} Martin, G., Kaviraj, S., Volonteri, M., et al.\ 2018, \mnras, 476, 2801.  

\bibitem[Martinet (1995)]{Mart95} Martinet L., 1995, FCPh, 15, 341. 

\bibitem[Merloni A. et al. (2010)]{mer10} Merloni A. et al., 2010, ApJ, 708, 137.

\bibitem[Mesa et al. (2014)]{mesa} Mesa V., Duplancic F., Alonso S., Coldwell G., Lambas D. G., 2014 MNRAS, 438, 1784.

\bibitem[Mihos \& Hernquist (1996)]{miher96} Mihos J. C. \& Hernquist L., 1996, ApJ, 464, 641. 

\bibitem[Moetazedian \etal (2017)]{Moet17} Moetazedian R., Polyachenko E.V., Berczik P., Just A., 2017, A\&A, 604, 95.

\bibitem[Mulchaey \etal (1994)]{mul94} Mulchaey, J. S., \etal, 1994, \apj, 436, 586.  

\bibitem[Nair \& Abraham (2010)]{na10} Nair, P. B., \& Abraham, R. G. 2010, ApJS, 186, 427.

\bibitem[Noguchi (1987)]{nogu87} Noguchi M., 1987, MNRAS, 228, 635.

\bibitem[Oh, Oh \& Yi (2012)]{oh12} Oh S., Oh K. \& Yi S. K., 2012, \apjs, 198, 40. 

\bibitem[Patton et al.(2016)]{patton16} Patton, D.~R., Qamar, F.~D., Ellison, S.~L., et al.\ 2016, \mnras, 461, 2589.

\bibitem[Pettitt \& Wadsley(2018)]{pet18} Pettitt, A.~R., \& Wadsley, J.~W.\ 2018, \mnras, 474, 5645.

\bibitem[Ramos Almeida \etal (2011)]{Ramos11} Ramos Almeida C., Tadhunter C. N., Inskip K. J., Morganti R., Holt J., Dicken D., 2011, MNRAS, 410, 1550.

\bibitem[Rees (1984)]{Rees84} Rees M.J., 1984, \araa, 22, 471. 

\bibitem[Reyes \etal (2008)]{roy08} Reyes R.,
Zakamska N., Strauss M. A., Green J., Krolik J. H., Shen Y., Richards G. T., Anderson S. F., Schneider D. P., 2008, AJ, 136, 2373.

\bibitem[Robichaud \etal (2017)]{robi17} Robichaud F., Williamson D., Martel H., Kawata D., Ellison S., 2017, MNRAS, 469, 3722.

\bibitem[Robotham et al.(2014)]{gama} Robotham, A.~S.~G., Driver, S.~P., Davies, L.~J.~M., et al.\ 2014, \mnras, 444, 3986.

\bibitem[Sabater \etal (2013)]{sab13} Sabater J., Best P. N., Argudo-Fernandez M., 2013, MNRAS, 430, 638.  

\bibitem[Sabater \etal (2015)]{sab15} Sabater J., Best P. N., Heckman, T. M., 2015, MNRAS, 447, 110.  

\bibitem[Sanders \etal (1988)]{sand88} Sanders, D.B., Soifer, B.T., Elias, J.H., Madore, B.F., Matthews, K., Neugebauer, G. \& Scoville, N., 1988, ApJ, 325, 74.

\bibitem[Shlosman, Begelman \& Frank (1990)]{SBF90} Shlosman I., Begelman M.C.\& Frank J., 1990, \nature, 345, 679. 

\bibitem[Shlosman, Frank \& Begelman (1989)]{SBF89} Shlosman I, Frank J., Begelman M. C., 1989, Natur, 338, 45.

\bibitem[Silverman (2011)]{sil11}  Silverman J. D. et al., 2011, ApJ, 743, 2. 

\bibitem[Simmons et al.(2017)]{sim17} Simmons, B.~D., Smethurst, R.~J., \& Lintott, C.\ 2017, \mnras, 470, 1559.

\bibitem[Smolcic, V., (2009)]{smol09}
Smolcic V., 2009, ApJ, 699, L43.

\bibitem[Strauss \etal (2002)]{strauss02} Strauss \etal 2002, \aj, 124, 1810. 

\bibitem[Tasse \etal (2008)]{ta08}
Tasse C., Best P. N., Rottgering H., Le Borgne D., 2008, A\&A, 490, 893.

\bibitem[Toomre \& Toomre (1972)]{tt72} Toomre A. \& Toomre J., 1972, ApJ, 178, 623.  

\bibitem[Tremonti \etal (2004)]{tremonti04} Tremonti C., Heckman T. M., Kauffmann, G. \etal, 2004, \apj, 613, 898.

\bibitem[Urrutia \etal (2008)]{urr08} Urrutia T., Lacy M., Becker R. H., 2008, ApJ, 674, 80. 

\bibitem[Urrutia \etal (2012)]{urr12} Urrutia T., Lacy M., Spoon H., Glikman E., Petric A., Schulz B., 2012, ApJ, 757, 125.

\bibitem[Vera \etal (2016)]{vera16} Vera, M., Alonso S., Colwell G., 2016, A\&A, 595, 63.

\bibitem[Zakamska et al.(2003)]{zaka03} Zakamska N., Strauss M. A., Krolik J. H., Collinge M., et al.\ 2003, \aj, 126, 2125. 

\bibitem[Zana et al.(2018)]{zana18} Zana, T., Dotti, M., Capelo, P.~R., et al.\ 2018, \mnras, 473, 2608. 

\bibitem[Zhou \etal (2015)]{zhou15} Zhou Z., Cao C., Wu H., 2015 \aj, 149, 1.


\end{thebibliography}
\end{document}